\documentclass[10pt,showpacs,amsmath,amssymb,floatfix,superscriptaddress,longbibliography,twocolumn,eqsecnum]{revtex4-2}
\usepackage{amsmath,amsthm,amsfonts,amssymb,bm}
\usepackage{amssymb}
\usepackage{graphicx}
\usepackage{subfigure}
\usepackage{overpic}
\usepackage{multirow}
\usepackage{bm}
\usepackage[usenames,dvipsnames,svgnames,table]{xcolor}
\usepackage{tcolorbox}
\counterwithout{equation}{section}
\counterwithout{figure}{section}
\usepackage[unicode=true,
            pdfusetitle,
            bookmarks=true,
            bookmarksnumbered=false,
            bookmarksopen=false,
            breaklinks=true,
            pdfborder={0 0 0},
            backref=false,
            colorlinks=true,
            hypertexnames=false]{hyperref}
\hypersetup{linkcolor=NavyBlue,urlcolor=NavyBlue,citecolor=NavyBlue}

\renewcommand{\arraystretch}{1.5}
\begin{document}

\title{Study of multiple degrees of freedom entanglement in optical fiber}
\author{Huimin Zhang}
\affiliation{School of Sciences, Hangzhou Dianzi University, Hangzhou 310018, China}
\author{Chaoying Zhao}
\email{zchy49@163.com}
\affiliation{School of Sciences, Hangzhou Dianzi University, Hangzhou 310018, China}
\affiliation{State Key Laboratory of Quantum Optics and Quantum Optics Devices, Institute of Opto-Electronics, Shanxi University, Taiyuan 030006, China}
 
\begin{abstract}
The orbital angular momentum (OAM) has attracted widespread attention due to its ability to carry information in multiple dimensions. However, a high-dimensional entanglement carrying OAM can be affected by environment and undergoes decoherence. Ensuring the stability and high fidelity of entangled states after transmission is a crucial part of quantum communication. How to control the entangled states are essential. In this paper, we produce the polarization entangled photon pairs by type I BBO crystals by means of spontaneous parametric down-conversion (SPDC), we achieve the polarization-OAM hybrid entangled states by q-plate (QP) by means of manipulating the multi-degrees of freedom of the quantum state after passing through the APD communication channel. The polarization entangled photon pairs have the characteristics of OAM. We use polarization degree of freedom to modulate OAM degree of freedom, our polarization-OAM hybrid entangled states  can slow down the reduction of the fidelity in the during of transmission process. Our quantum states exhibit a superior level of fidelity contrast with the conventional situation. This method will provide a theoretical guidance for improving the transmission fidelity of OAM states in fiber.
\end{abstract}  

\maketitle

\section{Introduction}
 The quantum information technology has led to revolutionary changes in some fields such as communication\cite{20-year,future,hd}, computing and cryptography\cite{jiami1,jiami2}.The concept of fidelity stands as a pivotal yardstick for gauging the quality of quantum information transmission.In the realm of quantum communication, the reliable conveyance of information hinges upon the accurate transmission of quantum states. However, owing to the sensitive characteristics of quantum systems and their susceptibility to environmental noise\cite{time-varying,noise1}, the practice of channeling quantum states through optical fibers\cite{1,2}, photonic crystal fibers\cite{23}, fiber Bragg gratings\cite{26,28,12}, and hollow-core fibers\cite{11} have become prevalent. However, in the transmission process, optical fibers are subject to phase shift\cite{14}, dispersion\cite{28,11} losses and inherent losses, etc. Therefore, how to realize high-capacity quantum state transmission as well as obtaining quantum states with higher fidelity has attract much attention.
 In the face of these problems, many efforts have been made. Researches have expand the degrees of freedom within quantum states to enhance their information-carrying potential, such as frequency degrees of freedom\cite{5}, path degrees of freedom\cite{6}, orbital angular momentum (OAM) degrees of freedom\cite{3,4}, and other degrees of freedom\cite{7,8,9,10}, among which, the OAM degrees of freedom play a huge role in increasing the capacity of quantum states due to their infinite dimensional properties.  The establishment of entanglement among diverse degrees of freedom can be realized through four-wave mixing\cite{on-chip}, spontaneous parametric down-conversion(SPDC)\cite{High-performance}, and the deployment of quantum memories\cite{Experimental}.    Consequently, strategies aimed at preserving the high fidelity of quantum state transmission through optical fibers take on paramount significance. Researches have put forward to using entanglement distillation\cite{distillation} technology can improve the fidelity of noisy entangled states. Researches also have added the phase to resist the effect of the phase shift generated by the optical fiber to improve the fidelity after transmission\cite{14}.  
This article aims to delve into the intricate relationship between quantum states and fidelity, while spotlighting the pivotal role quantum states play within optical fiber transmission. we propose a method to modulate the quantum state by modulating the polarization degree of freedom to affect the other degrees of freedom. By methodically manipulating quantum states to optimize fidelity, we embark on a comprehensive exploration that includes discussions on noise channels and the specific intricacies encountered within optical fiber environments. We find that the modulated entangled state can have a better resistance to noise. Our research result can provide a favorable help to guide how to achieve a high-fidelity quantum state transmission.

\section{The multiple
degree of freedom entanglement generation via spontaneous parametric down conversion (SPDC) processes in BBO crystal}
A Gaussian pump beam with a wavelength of $\lambda=1.55\mu m$ goes incident on a polarization beam splitter (PBS) via a mode-matching lens and a high birefringent Beta-Barium Borate (BBO) plate, and then passes through two optical paths including a $q$-plate\cite{1,oam-beam} and two quarter-wave plate (QWP). Here, PBS is used in order to ensure the polarization alignment for maximizing the efficiency of phase matching in BBO plate. The entanglement can be manipulated by modulating the optical angle sensor in a digitally enhanced homodyne interferometer in experiments. The measurement sensitivity of the optical angle sensor can be influenced by the BBO plate and the environmental condition. We utilize the spontaneous parametric down-conversion (SPDC) processes to generate signal-idle photon pairs\cite{SPDC,spdc1,spdc2}. The frequency degree of freedom refer to the frequency difference between the generated signal photons and idle photons, which can be controlled. 0 and 1 can be looked as a quantum representation of the degree of freedom, defining the frequency of one photon as 0 and the other as 1, and forming an entanglement after passing through the SPDC.The OAM degree of freedom can be carried after passing through the $q$-plate. The reason for putting only one of the two optical paths generated is that the two optical paths are compared and then the effect of the noise channel environment on the photons is compared, as well as the remaining fidelity. The entanglement formed by the upper optical path and the lower optical path is consistent, one of entanglements passes through the APD (amplitude and phase damping) channel, another passes through a single mode fiber(SMF), which can filter the topological charge non-zero photons and purify the polarization entangled photon pairs. We choose the radius of the fiber to be $9.5\mu m$, when the value becomes larger, there are more modes, the coupling becomes more complex, when the value becomes smaller, which may be produce a different modes transfer.

After described the photon pair source generation, we then went further to demonstrate the generation of multiple degrees of freedom entanglement. 

Firstly,  we carry out a theoretical treatment for the quantification of polarization entanglement for the bi-photon state:
\begin{gather}
|\Psi_{pol}\rangle =\alpha|\psi^\dagger(\theta,\phi)\rangle|\psi^\dagger(\theta,\phi)\rangle
+\beta|\psi^-(\theta,\phi)\rangle|\psi^-(\theta,\phi)\rangle\\
|\psi^\dagger(\theta,\phi)\rangle =cos(\frac{\theta}{2})|0\rangle+e^{i\phi}sin(\frac{\theta}{2})|1\rangle \\
|\psi^-(\theta,\phi)\rangle=sin(\frac{\theta}{2})|0\rangle-e^{i\phi}cos(\frac{\theta}{2})|1\rangle
\end{gather}
 which can be continuously varied across a wide range of separable and entangled states, depending on the parameters $\alpha$ and $\beta$, and $\alpha^2+\beta^2=1$.  The normalized parameter $\alpha$ can be controlled by changing the nonlinear  birefringent BBO crystal\cite{spdc1}, the value of $\alpha$ maybe obtained by a seventy-two degrees rotation of a QWP in one of the optical paths. $\theta$ is the Angle of phase-matching between the pump and the optics axis of BBO crystal. Through the distribution of the phase matching angles, we can obtain the required type of entanglement.$\phi$ is the phase difference between the two optical path. $|\psi(\theta,\phi)\rangle$ is the state that already select the values of $\theta$ and $\phi$. The polarization entanglement state takes this form mainly due to the value of $\theta$ and $\phi$ and the form of the state composed of qubits, and the value of $\theta$ and $\phi[0,pi]$ is selected because of its periodicity. This value determines the form of entanglement chosen in our paper. $|\psi^\dagger(\theta,\phi)\rangle$ and $|\psi^-(\theta,\phi)\rangle$ are the expressions of arbitrary basis of a single qubit(we use the superscript $'+'$ and $'-'$ to distinguish the different basis).  The probabilities of the entangled states generated through the SPDC process is $cos(\theta/2)$ and $sin(\theta/2)$, respectively. 

Secondly, the initial Guassian state defined as $|0\rangle$ in order to more conveniently describe the change of photons after passing through the device, and the subsequent $|0\rangle$ and  $|1\rangle$ represents the OAM entangled state with topological charge $l=1$ and  $l=-1$, respectively. The entangled state comprises two photons. We neglect the fundamental mode, the OAM entangled states $l=1$ and $l=-1$ are analogous to the cases of logical $0$ and $1$, respectively. Let's consider the evolution of a single photon as follows: 
\vspace{-0.15em}
\begin{gather}
|0\rangle|H\rangle\stackrel{QWP}{\longrightarrow}|0\rangle|R\rangle\overset{q-plate}{\underset{q=l/2}\longrightarrow}|l\rangle|L\rangle\stackrel{QWP}{\longrightarrow}|l\rangle|H\rangle\\
|0\rangle|V\rangle\stackrel{QWP}{\longrightarrow}|0\rangle|L\rangle\overset{q-plate}{\underset{q=l/2}\longrightarrow}|-l\rangle|R\rangle\stackrel{QWP}{\longrightarrow}|-l\rangle|V\rangle
\end{gather}
where $|L\rangle$ represents the left-handed circularly polarized light, and $|R\rangle$ represents the right-handed circularly polarized light. The $q$-plate has a order of $l/2$, and the generated OAM light with the topology change $l$. This allowed us to transform the initial polarization entangled state $|\Psi_{pol}\rangle$ into the state $|\Psi\rangle$. 

Thirdly, a quantum system is often affected by an external environment, the dissipation is a irreversible process, which will result in a decoherence phenomenon. We will adopt the amplitude and dephase channel method\cite{time-varying} for simulating the decoherence process. In this paper, we set entangled beams passing through a hybrid noise channel optical fiber, which composed of the amplitude damping (AD) noise channel and the phase damping (PD) noise channel. We take the proportions of both the AD noise channels and the PD noise channels $50\%$, namely the influence of the two noise channels is evenly balanced. $T_1$ is the relaxation time of the AD noise channel, $t_1$ is the transmission time through the AD noise channel. $T_2$ is the dephasing time of the PD noise channel\cite{time-varying}, and $t_2$ is the transmission time through the PD noise channel. The static noise channel need consider dispersion, a low aberration coefficient ($1\%$) will make the dispersion becomes small, so the effect of dispersion should not be considered. We set $t_1=t_2=100\mu s$\cite{time-varying}, we can see the advantages under this condition more clearly. When the amplitude and dephase damping noise acts on any arbitrary quantum state $\rho$, there is a certain probability of inverting qubits\cite{time-varying}, $\varepsilon (\rho)=|\psi\rangle\langle\psi|$ denotes the quantum states unaffected by noise. Therefore, the Krause operators $E_0$, $E_1$, $E_2$ in the APD noise channel have the following relationships:
\begin{gather}
\rho=|\Psi\rangle\langle\Psi|\\
\varepsilon (\rho)=\sum^{2}_{m,n=0} \widehat{E}_m \otimes \widehat{E}_n|\Psi\rangle \langle\Psi|\otimes\widehat{E}_m^{\dagger} \otimes \widehat{E}_n^{\dagger}\\
E_0=
\begin{bmatrix}
    1 & 0\\
    0 & \sqrt{1-\gamma_d -(1-\gamma_d )\gamma_s }
\end{bmatrix}\\
E_1=
\begin{bmatrix}
    0 & \sqrt{\gamma_d }\\
    0 & 0
\end{bmatrix}\\
E_2=
\begin{bmatrix}
    0 & 0\\
    0 & \sqrt{(1-\gamma_d )\gamma_s }
\end{bmatrix}
\end{gather}
where $\gamma_d $ represents the damping noise and $ \gamma_s $ represents the scattering noise. 

\begin{figure}
\begin{minipage}[]{0.45\linewidth}
\centering
\begin{overpic}[scale=0.23]{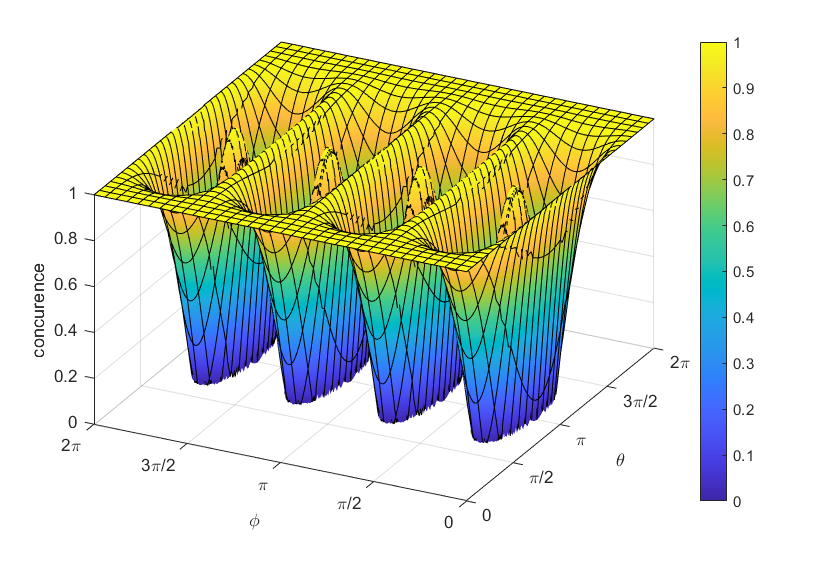}
\put(3,70){\textbf{(a)}}
\end{overpic}
\label{1a}
\end{minipage}%
\begin{minipage}[]{0.45\linewidth}
\centering
\begin{overpic}[scale=0.23]{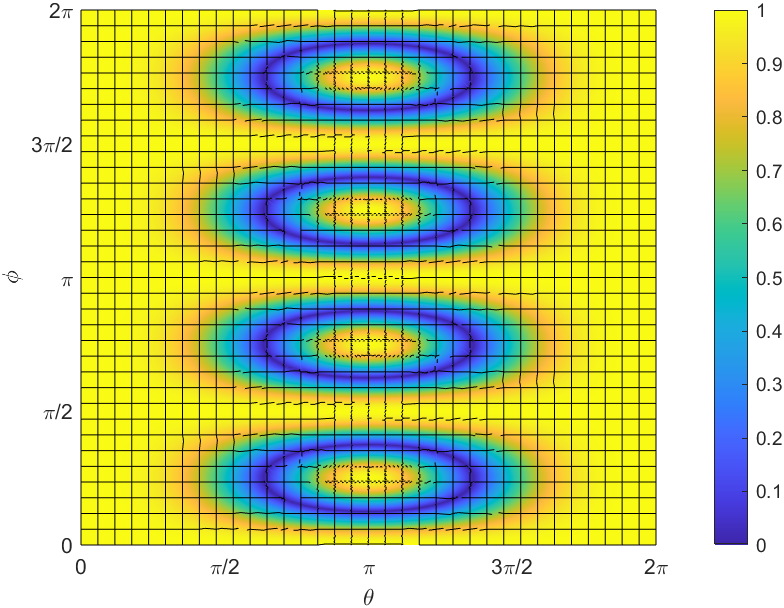}
\put(3,70){\textbf{(b)}}
\end{overpic}
\label{1b}
\end{minipage}%
\caption{(a)Variation of a concurrence parameter ($C(\Psi)$)
) with the Angle of phase-matching ($\theta$) and the phase difference ($\phi$)
) in 3-D plot. The maximum value of $C(\Psi)$ ($C(\Psi)=1$) can be attained when $\theta=n\pi/4$, $\phi=\pi$.(b)The projection of(a) on $\theta$ and $\phi$ two dimensional images.}
\label{1}
\end{figure}

\begin{figure}
\subfigure{
\centering
\begin{overpic}[scale=0.21]{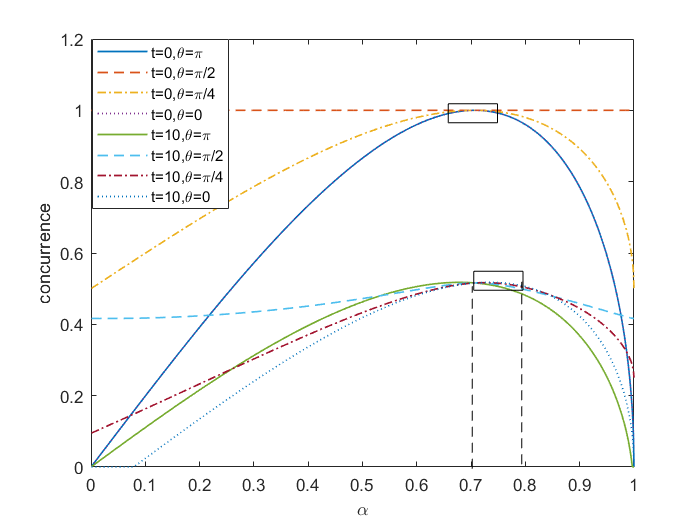}
\put(3,73){\textbf{(a)}}
\end{overpic}
\label{3a}}
\subfigure{
\centering
\begin{overpic}[scale=0.21]{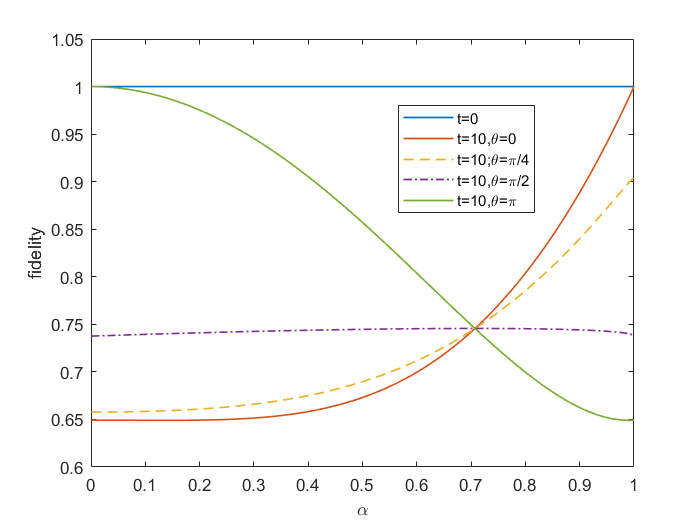}
\put(3,73){\textbf{(b)}}
\end{overpic}
\label{3b}}
\subfigure{
\centering
\begin{overpic}[scale=0.21]{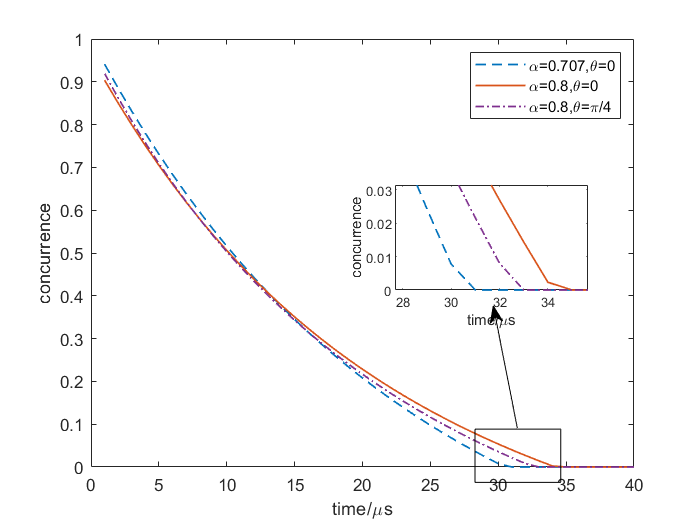}
\put(3,73){\textbf{(c)}}
\end{overpic}
\label{3c}}
\subfigure{
\centering
\begin{overpic}[scale=0.21]{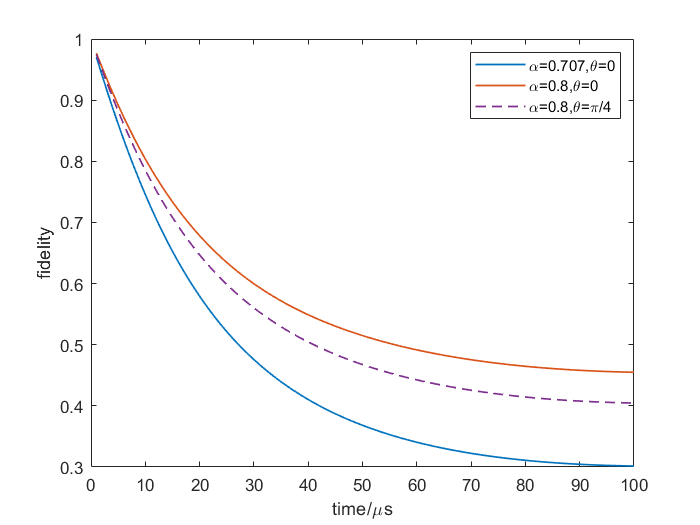}
\put(3,73){\textbf{(d)}}
\end{overpic}
\label{3d}}
\caption{4 qubits hybrid entangled when $\phi=\pi$. (a)Variation of concurrence with $\alpha$.(b)Variation of fidelity with $\alpha$. The relationship between "concurrence and fidelity" and variation of $\alpha$ with $\theta$ Angles. The region enclosed within the boxed area represents the range of select-able "concurrence" values (without considering $\theta=\pi$ case).(c)Variations of concurrence with $t$.(d)Variations of fidelity with $t$ under different $\alpha$ and $\theta$ Angles.}
\label{3}
\end{figure}

Lastly, there are various measure methods were proposed in order to quantify the entanglement, the most famous is concurrence. With the help of quantum state tomography, concurrence can reconstitute a high-dimensional entanglement\cite{15}. In this paper, we can calculate concurrence of any pure state $\Psi$ by using the following relations: 
\begin{gather}
C(\Psi)=|\langle\psi|\widetilde{\psi}\rangle|^2\\
|\widetilde{\psi}\rangle=\sigma_y^{\otimes n}|\psi\rangle
\end{gather}
where $\sigma_y$ is the Pauli matrix, and $n$ represents the number of qubits. For the sake of computational convenience, with the increasing of $\theta$ and $\phi$, the concurrence maintains a noise-free situation(see Fig.\ref{1}).
\begin{figure}
\subfigure{
\centering
\begin{overpic}[scale=0.20]{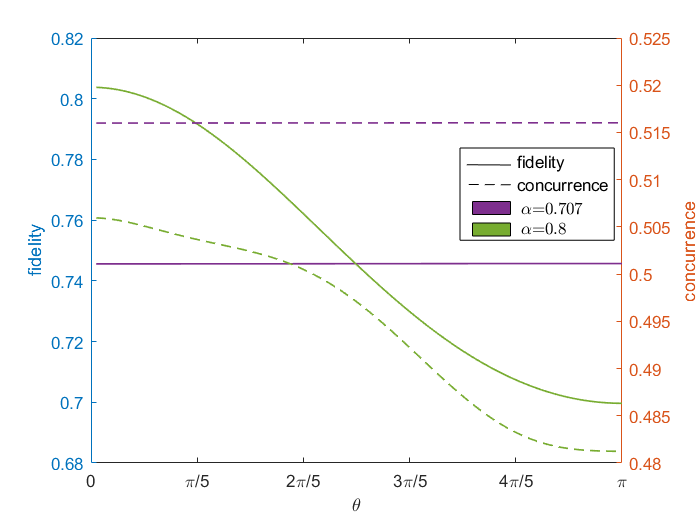}
\put(3,73){\textbf{(a)}}
\end{overpic}
\label{2a}}
\subfigure{
\centering
\begin{overpic}[scale=0.20]{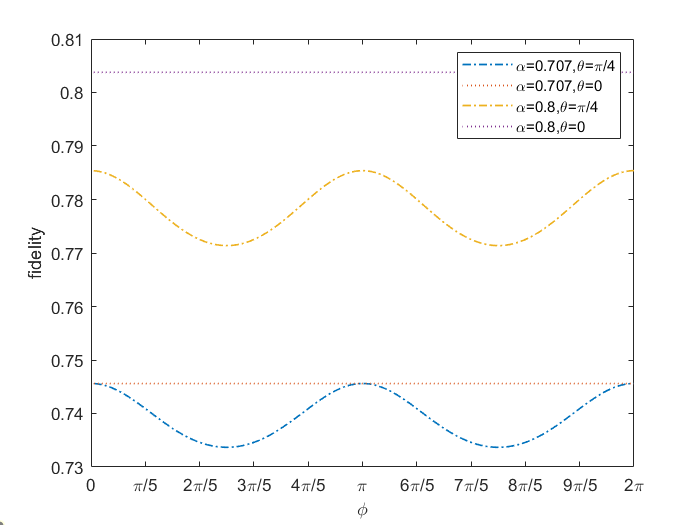}
\put(3,73){\textbf{(b)}}
\end{overpic}
\label{2b}}
\caption{(a)Variation of fidelity and concurrence with $\theta$, the purple solid line corresponds to $\alpha=0.707$ and the green solid line corresponds to $\alpha=0.8$. (b)Variation of fidelity with $\phi$. The solid line represents $\theta=0$ case. The dashed line represents $\theta=\pi/4$ case. When $\theta=0, \pi/4$, the concurrence and fidelity are better than that of the $\theta=\pi/2,\pi$ situation}.
\label{2}
\end{figure}
When we choice $\phi=\pi$, $t=0$, we find that $\alpha=1/\sqrt{2}=0.707$ is the noise-free situation. If we consider the influence of noise, $\alpha$ is no longer a value of constant, but different values of changes over time. We demonstrate the changes in fidelity and concurrence under different values of $\alpha$.(see Fig.\ref{3}). When $\theta$ changes over time, by numerical simulating, we find that $\alpha\in(0.65,0.8)$ states can get a better results from Fig.\ref{3a}. Form Fig.\ref{2a}, the value of $\theta=\pi$ is relatively small, therefore, we need not take into account this situation. In the following calculations, firstly, we choose $\alpha=0.65,0.707,0.8$, respectively. By comparing the conclusions of Fig.\ref{3c}-\ref{3d}, we can easily find that $\theta=0, \alpha=0.8$ state can yields a longer survival time and a higher fidelity and concurrence with the time increasing. Secondly, we continue to investigate $\phi$ changes over time when given $\theta=0, \alpha=0.8$. Referring to Fig.\ref{2b}, we find that the values of $\phi$ have no impact on the overall outcome. However, it is still necessary to consider the impact of $\phi$ on the overall outcome when $\theta\neq 0, \alpha=0.8$. Thus, the output states from the entangled qubits are:
\begin{gather}
|\Psi_{final}\rangle=0.8|HH\rangle\otimes|l,l\rangle+0.6|VV\rangle\otimes|-l,-l\rangle
\end{gather}
The scenarios described above did not account for the entanglement between frequencies that accompanies the process of SPDC-generated entanglement. Therefore, the subsequent discussion will focus on the frequency degree of freedom.
\begin{gather}
|\Psi\rangle=|\Psi_{final}\rangle\otimes|\Psi_{ fre}\rangle\\ |\Psi_{fre}\rangle=|10\rangle+|01\rangle
\end{gather}
Upon introducing the frequency degree of freedom, the overall trend of the states remains consistent with the previous scenario in the absence of noise. Therefore, we continue to use the same values of $\theta$ and $\phi$ as 4 qubits case, thus, we can determine the range of value of $\alpha$.

\begin{table*}
\label{34d}Table.1 The improvement of fidelity between $\alpha=0.8$ and $\alpha=0.707$. 
\begin{ruledtabular}
\begin{tabular}{ccccccc}
 &time/$\mu$s&10&30&50&80&100\\ \hline
 \multirow{2.1}{*}{$4-qubits$} 
&$\alpha=0.707$&0.7456&0.4557&0.3682&0.3109&0.3016\\
&$\alpha=0.8$&0.8038&0.5997&0.5149&0.4647&0.4549\\ \hline
&Fidelity improve ratio &7.81$\%$&31.60$\%$&39.84$\%$	&49.47$\%$&50.83$\%$ \\ \hline
\multirow{2.1}{*}{$6-qubits$} 
&$\alpha=0.707$&0.6347&0.2876&0.1601&0.0976&0.0769\\& $\alpha=0.8$&0.6843&0.3626&0.2291&0.1439&0.116\\ \hline
&Fidelity improve ratio &7.81$\%$&26.08$\%$	&43.10$\%$&47.44$\%$&50.85$\%$
\end{tabular}
\end{ruledtabular}
\end{table*}

\begin{table*}
\label{5}Table.2 The improvement of fidelity between $\alpha=0.8$ and $\alpha=0.707$ with different $l$.
\begin{ruledtabular}
\begin{tabular}{ccccccccc}
 &time/$\mu$s&10&30&50&80&100&200&300\\ \hline
 \multirow{2.1}{*}{ $l$=4} 
 &$\alpha=0.707$&0.8806&0.6828&0.5552&0.4507&0.4123&0.3327&0.2913
\\
 &$\alpha=0.8$&0.8956&0.7177&0.5988&0.498&0.4598&0.3773&0.3342
\\ \hline
&Fidelity improve ratio & 1.70$\%$&5.11$\%$&7.85$\%$&10.49$\%$&11.52$\%$&13.41$\%$&14.73$\%$
 \\ \hline
\multirow{2.1}{*}{ $l$=8} 
 &$\alpha=0.707$&0.8351&0.5819&0.4335&0.3218&0.2831&0.2071&0.1709
\\
& $\alpha=0.8$&0.8493&0.6117&0.4675&0.3555&0.3157&0.2349&0.1961
\\ \hline
&Fidelity improve ratio &1.70$\%$&5.12$\%$&7.84$\%$&10.47$\%$&11.52$\%$&13.42$\%$&14.75$\%$ \\ \hline
\multirow{2.1}{*}{ $l$=16} 
 &$\alpha=0.707$&0.7919&0.4959&0.3384&0.2297&0.1944&0.129&0.1003
\\
 &$\alpha=0.8$&0.8054&0.5213&0.365&0.2598&0.2167&0.1463&0.115
\\ \hline
&Fidelity improve ratio &1.70$\%$&5.12$\%$&7.86$\%$&13.10$\%$&11.47$\%$&13.41$\%$&14.66$\%$
\end{tabular}
\end{ruledtabular}
\end{table*}
 In this paper, the situation that the fidelity in the case of $ t=10\mu s$  for 6 qubits in the noise channel like the situation that the fidelity can reaches to $ 0.71 $ after transmission of $ 1 km $ for distribution \cite{14}, which use both OAM and frequency entanglement. If we set the propagation speed in optical fiber equal to $0.2 ns/m$, the transmission distance should be $ 2 km $. The transmission distance is converted to $t=5$, if adopt our method, fidelity will be higher.
\begin{figure}
\subfigure{
\centering
\begin{overpic}[scale=0.21]{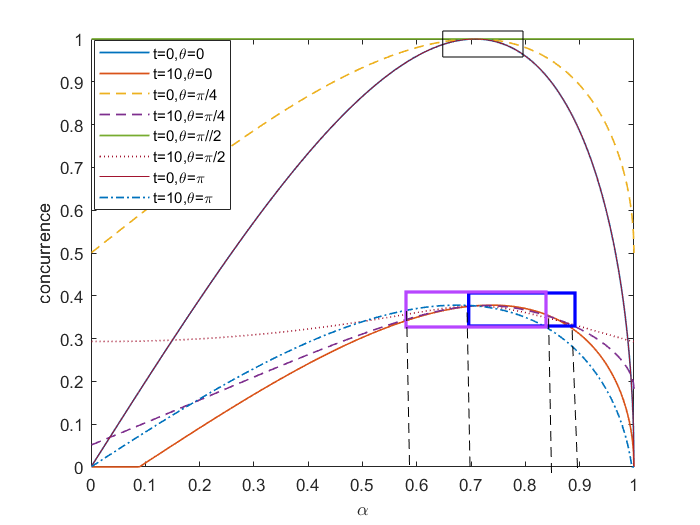}
\put(3,73){\textbf{(a)}}
\end{overpic}
\label{4a}}
\subfigure{
\centering
\begin{overpic}[scale=0.21]{aft10}
\put(3,73){\textbf{(b)}}
\end{overpic}
\label{4b}}
\subfigure{
\centering
\begin{overpic}[scale=0.21]{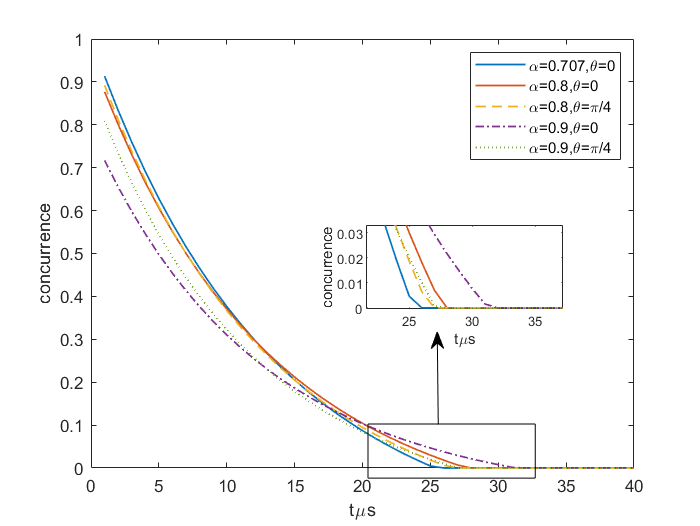}
\put(3,73){\textbf{(c)}}
\end{overpic}
\label{4c}}
\subfigure{
\centering
\begin{overpic}[scale=0.21]{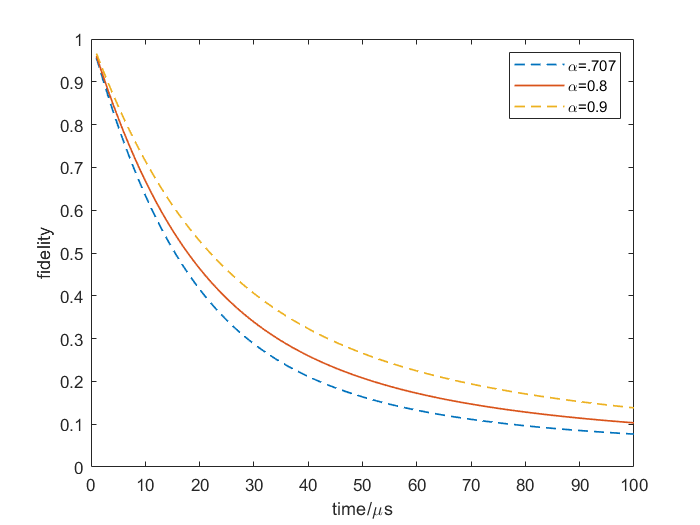}
\put(3,73){\textbf{(d)}}
\end{overpic}
\label{4d}}
\centering
\caption{6 qubits hyper- entangled when $\phi=\pi$.(a)Variation of concurrence with $\alpha$. The lower left region  is the vortex area of $\theta$=$\pi/2$ and $\theta$=$\pi$, while the lower-right region corresponds to the vortex area of $\theta$=$\pi/4$ and $\theta=0$. (b)Variation of fidelity with $\alpha$. (c)Variation of concurrence with $\alpha$.(d)Variation of fidelity with $\alpha$.}
\label{4}
\end{figure}
For 6 qubits cases, we consider three degrees of freedom, including  polarization, frequency, and OAM, whereas the OAM degree of freedom can realize a high-dimensional entanglement, the polarization degree of freedom serves mainly to modulate the quantum state.
Due to the linear polarization direction in actually fiber will changes with the azimuthal angle of the change \cite{mao_complex_2019}, which will lead to a instability in transmission process. Therefore, we should convert the linear polarization into circular polarization before put it into the optical fiber, and the circular polarization and the OAM degrees of freedom can form a stable OAM mode.

\begin{table*}
\renewcommand{\arraystretch}{1.5}
\label{5}Table.3 Comparison of seven schemes.
\begin{ruledtabular}
\begin{tabular}{cccccc}
 &Entanglement type&Environmental&Modulation method&Conversion Efficiency&Fidelity
\\ \hline
2015\cite{spdc-based}&Polarization&-&-&-&99$\%$\\
2018\cite{18-qubit}&Polarization-OAM&-&-&-&70.8$\%$\\
2018\cite{coherent}&OAM&-&SLM(phase)&-&85.81$\%$\\
2020\cite{14}&OAM&fiber(1km)&SLM(phase)&-&71$\%$\\
2020\cite{distillation}&Polarization-OAM&AD channel\footnote[1]{Amplitude Damping Channel}&Distillation&-&60$\%$\footnote[2]{we choose Two-photons case, and $T_1$=$T_2$, $\lambda=0.8$.}\\ 
2021\cite{17}&SAM-OAM&-&fiber& 90$\%$&-\\
\multirow{2}{*}{ Our work} 
&Polarization-OAM&APD Channel\footnote[3]{Amplitude and Phase Damping Channel}(2km\footnote[4]{we seleted 4 qubits cases and $t=10\mu s$, consider the speed of light as $0.2ns/m$. })&QWP&-&80.38$\%$ \\
&Polarization-OAM&APD Channel(2km\footnote[5]{we seleted 6 qubits cases and $t=10\mu s$, consider the speed of light as $0.2ns/m$. })&QWP&-&68.43$\%$
\end{tabular}
\end{ruledtabular}
\end{table*}

\begin{figure}
\centering
\begin{overpic}[scale=0.3]{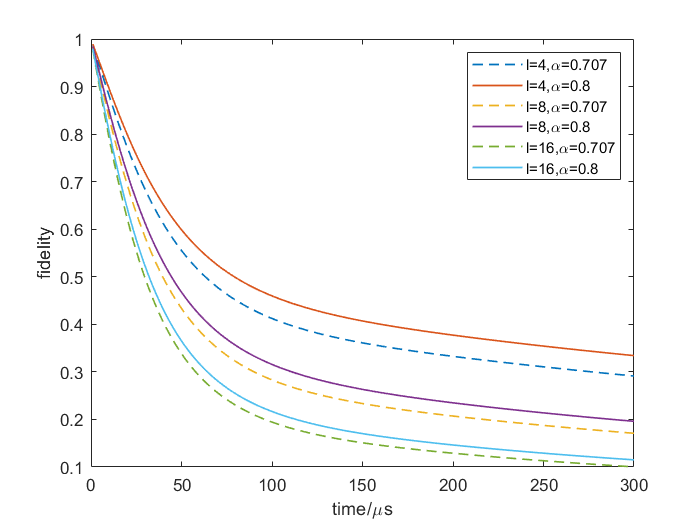}
\put(3,100){\textbf{}}
\end{overpic}
\caption{Variation of fidelity with topological charge $l$ and $\alpha$. The red solid line, the purple solid line and the blue solid line corresponds to $\alpha=0.8$ and $l=4,8,16$, respectively. The blue dashed line, the orange dashed line and the green dashed line corresponds to $\alpha=0.707$ and $l=4,8,16$, respectively.}
\label{8}
\end{figure}

The inclusion of the frequency degree of freedom is reflected in Fig.(\ref{4a}) graph, where the range of value of $\alpha$ has transitions from (0.65-0.8) to (0.7-0.9), undergone a tiny right shifts. Consequently, in the case of 6 qubits, we set $\alpha=0.9$. However, upon closer examination (of see Fig.(\ref{4d}))we note that $\alpha=0.9$ has a higher fidelity over time compared to another two cases. However, $\alpha=0.9$ has a notably low-lying initial entanglement levels, so $\alpha=0.9$ isn't a favorable choice. Considering both Fig.(\ref{4}) and Fig.(\ref{5}), $\alpha=0.8$ is an optimal value.
 OAM mentioned above are solely considering the individual state $|l\rangle$, without taking all these states into account such as $|l-1\rangle$,...,$|1\rangle$. For example, if we take $l=4$, we should take into account $|4\rangle$, $|3\rangle$, $|2\rangle$ and $|1\rangle$ at the same time\cite{q-plate}. Therefore, it's necessary to discuss the OAM degree of freedom entanglement. Fig.\ref{8} unmistakably illustrates that the Fidelity of $\alpha=0.8$ consistently surpasses that of $\alpha=0.707$ case.

\begin{figure}
\subfigure{
\centering
\begin{overpic}[scale=0.21]{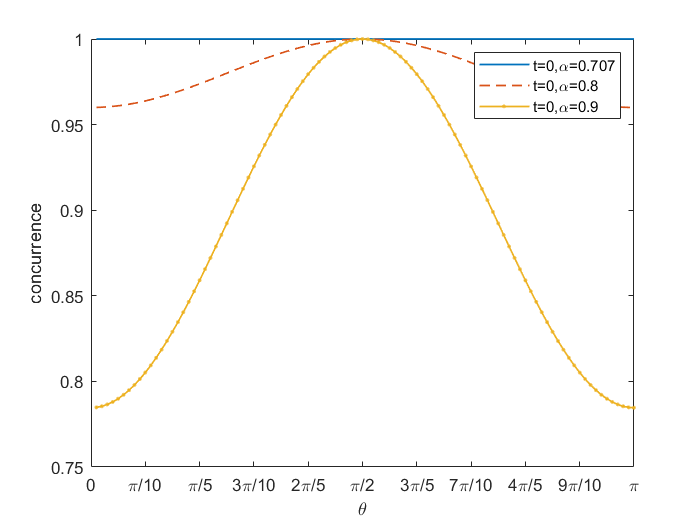}
\put(3,73){\textbf{(a)}}
\end{overpic}
\label{5a}}
\subfigure{
\centering
\begin{overpic}[scale=0.21]{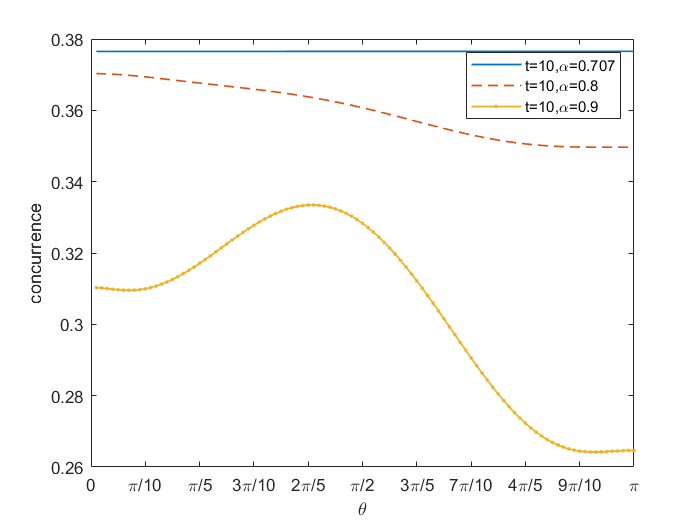}
\put(3,73){\textbf{(b)}}
\end{overpic}
\label{5b}}
\centering
\caption{Variation of concurrence  with $\alpha$ and $t$. (a)Variation of concurrence under noiseless conditions when $t=0$. (b)Variation of concurrence with $\theta$ when $t=10$.}
\label{5}
\end{figure}

We give out a preliminary estimate of the value of $\alpha$ is $0.8$, but it is not an optimal value in the presence of noise. The ideal value of $\alpha$ will varies depending on different actual conditions. Nevertheless, $\alpha$ with an approximate range of 0.8, which tends to improve the Fidelity of entanglement. Next, Fibre transport of the spatially entangled 4 qubits through the optical fibers and examine whether the value of $\alpha$ is applicable to our situation.

 \section{Entanglement evolution in optical fiber in the presence of noise}
We find that the linear polarization direction will changes with the azimuthal Angle\cite{mao_complex_2019}, the optical characteristics of the linearly polarized modes in optical fiber will result in an instability problem, we contemplate on a QWP in order to transforming the linearly polarized mode into a circularly polarized mode before entering the optical fiber. The circularly polarized mode with the OAM degrees of freedom can form a stable OAM mode. The prepared enantagled state may be affected by various factors such as dispersion\cite{28}, and noise\cite{29}. Therefore, the entangled state in the optical fiber can be expressed as \cite{17}:
\begin{gather}
\hat{U}_{fiber} = \int f(\omega_s,\omega_i)e^{i\beta_{fiber} z} dr\\
\beta_{fiber}=\frac{2\pi n_{eff}}{\lambda_{photon}}\\
|\Psi_{fiber}\rangle = \hat{U}_{fiber}|\Psi_{in}\rangle\notag\\
=\int f(\omega_s,\omega_i)d\omega_sd\omega_i(0.8e^{i\beta_1z}|1,1\rangle \notag\\
+0.6e^{i\beta_2z}|-1,-1\rangle)\otimes|\Psi_{fre}\rangle
\end{gather}
where $\alpha=0.8$, $\beta=0.6$. $z$ is the transmission distance in hybrid channel. The topological changes $l$ of OAM depends on the order of the $q$-plate and the circularly polarized state. $|-1,-1\rangle$ state generated by $q$-plate. $|\Psi_{in}\rangle$ is the initial state prepared outside the fiber, we set $m=1$, and the evolution operator $\hat{U}_{fiber}$ represents the influence of the fiber on the state. where $\beta$ is the propagation constant in the optical fiber, while $n_{eff}$ is the effective refractive index within the fiber. Typically, these values are associated with the optical fiber mode, OAM topology, and frequency, $\beta_1$ is the propagation constant when topology $l=1$, and $\beta_2$ is the propagation constant when topology $l=2$. $z$ is the transmission distance in the fiber, $\lambda_{photon}$ is the wavelength of photon. $\omega_s$ and $\omega_i$ is the frequency of signal and idle photon, respectively, $f(\omega_s,\omega_i)d\omega_sd\omega_i$ represents the radial distribution function, and $\int|f(\omega_s,\omega_i)|^2d\omega_sd\omega_i=1$.

We discuss hyper-entangled photon pairs in optical fibers\cite{10}, we calculate the tensor product of SAM and OAM degrees of freedom, which are generated directly through SPDC process. We can derived out the CHSH inequality (except the frequency DOF) by projecting the entanglement into sub-spaces, where $\hat{P}$ is the project operator, $X$, $Y$ and $Z$ are the Pauli matrices.
\begin{gather}
  |h_1\rangle=(|H\rangle+|v\rangle)/\sqrt{2},
    |v_1\rangle=(|H\rangle-|v\rangle)/\sqrt{2}\\\notag
    |h_2\rangle=(|m\rangle+|-m\rangle)/\sqrt{2},
    |v_2\rangle=(|m\rangle-|-m\rangle)/\sqrt{2}\\\notag
    \hat{P}_1=|h_1h_1\rangle\langle h_1h_1|\\\notag
    =(|HH\rangle+|VV\rangle+|HV\rangle+|VH\rangle)\\\notag
    (\langle HH|+\langle VV|+\langle HV|+\langle VH|)/4\\\notag
    =(|HH\rangle+|VV\rangle+|HV\rangle+|VH\rangle)(\langle HH|+\langle VV|)/4+...\\\notag
    \hat{P}_2=|v_1v_1\rangle\langle v_1v_1|,
    \hat{P}_3=|h_2h_2\rangle\langle h_2h_2|\\\notag
    \hat{P}_4=|v_2v_2\rangle\langle v_2v_2|,
    |\psi_p\rangle=\hat{P}|\Psi_{final}\rangle\\\notag
    \hat{A}_1=(X+Z)/\sqrt{2},\hat{A}_2=(X-Z)/\sqrt{2}\\\notag
    \hat{B}_1=X, \hat{B}_2=Z\\\notag
    \hat{a}_1=(X+Y)/\sqrt{2},\hat{a}_2=(X-Y)/\sqrt{2}\\\notag
    \hat{b}_1=X,\hat{b}_2=Y\\\notag
    \langle \hat{A}\hat{B}\rangle= \langle \psi_p|\hat{A}\hat{B}|\psi_p\rangle\\\notag
    CHSH_{pol}=\langle \hat{A}_1\hat{B}_1\rangle+\langle \hat{A}_1\hat{B}_2\rangle+\langle \hat{A}_2\hat{B}_1\rangle-\langle \hat{A}_2\hat{B}_2\rangle\\\notag
    CHSH_{OAM}=\langle a_1b_1\rangle+\langle a_1b_2\rangle+\langle a_2b_1\rangle-\langle a_2b_2\rangle\\\notag
    X=
    \begin{bmatrix}
    0 & 1\\
    1 & 0
\end{bmatrix}
        Y=
    \begin{bmatrix}
    0 & -i\\
    i & 0
\end{bmatrix}
        Z=
    \begin{bmatrix}
    1 & 0\\
    0 & -1
\end{bmatrix}
\end{gather}
By calculation, we can get the different project operations, $\hat{P}_1$ and $\hat{P}_2$ are the same, $\hat{P}_3$ and $\hat{P}_3$ are the same, and $CHSH_{pol}$=2.8284, it is the Maximum entanglement, $CHSH_{OAM}$=2.7153, it is not the Maximum entanglement due to the SAM entanglement with a coefficient. 

\begin{figure}
\centering
\subfigure{
\begin{overpic}[scale=0.21]{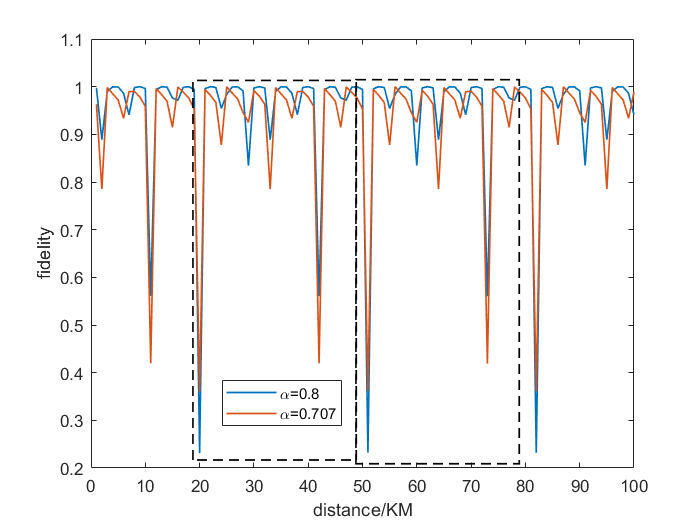}
\put(3,73){\textbf{}}
\end{overpic}}
\caption{Variation of fidelity with transmission distance}
\label{6}
\end{figure}

In this paper, we adopt standard single-mode fiber\cite{14}. In addition to the calculation of quantum noise mentioned above, we must consider the whole intrinsic fiber loss such as the absorption loss, the dispersion loss and the scattering loss caused by the fiber structural defects. In the following calculation, we take the whole intrinsic fiber loss equal to $0.36 dB/km$\cite{27}. Understanding the transmission process is crucial for investigate the evolution of entanglement, we introduce phase shifts factors when topological charge $l=1$($\theta_1=1.02\pi$, $\phi_1=0.98\pi$\cite{14}). Based on the effective refractive index $n_{neff}$ in the optical fiber, we find that between propagation mode $m=1$ and propagation mode $m=-1$ indication of a substantial degeneracy\cite{degeneracy}. Hence, we select the value of $n_1=n_{-1}=1.448$\cite{14}. The  pump wavelength $\lambda_{pump}=1550nm$. We investigate the fidelity changes with the transmission distance.(see Fig.\ref{6}).
From the two dotted boxes in Fig.\ref{6} we find that the Fidelity curves has a clear oscillation with stable periodic characteristics. Previous studies have reported that researcher can convert periodically spin angular momentum (SAM) to OAM in weakly-coupled few-mode fibers\cite{17}, if we furthermore convert this result to the fidelity, which is in agreement with periodic characteristic.
The entire state fidelity is
\begin{gather}
\begin{aligned}
|\Psi_{fiber}\rangle ={} & \int f(\omega_s,\omega_i)d\omega_sd\omega_i(0.8e^{i\beta_1z}e^{i\theta_1}|1,1\rangle\\
& + 0.6e^{i\beta_2z}e^{i\phi_1}|-1,-1\rangle)\otimes|\Psi_{fre}\rangle\\
\end{aligned}
\end{gather}
\vspace{-0.15em}
It is evident from Fig.(\ref{6}) that there isn't a substantial disparity between $\alpha=0.707$ and $\alpha=0.8$ at the peaks values of the Fidelity. Nevertheless, the distinct advantage lies in the fact that $\alpha=0.8$ has a broader range of favorable positions in comparison to that of $\alpha=0.707$ case. Subsequently, we shift the focus to examining the impact of optical fiber on the OAM degree of freedom after transmission has been completed. For this purpose, we use the two-measurement tomography method\cite{two-measurement} to reconstruct the output beam when $z=95km$. We can observe the changes in the OAM degree of freedom before and after the output. 
\begin{figure}
\subfigure{
\centering
\begin{overpic}[scale=0.21]{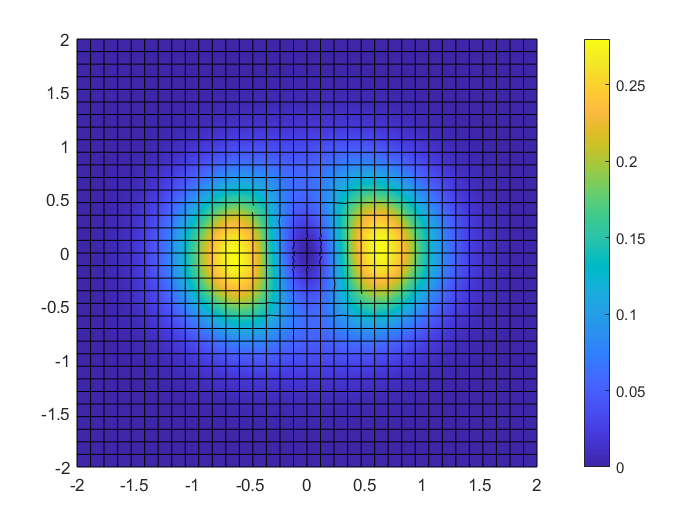}
\put(3,73){\textbf{(a)}}
\end{overpic}
\label{7a}}
\subfigure{
\centering
\begin{overpic}[scale=0.21]{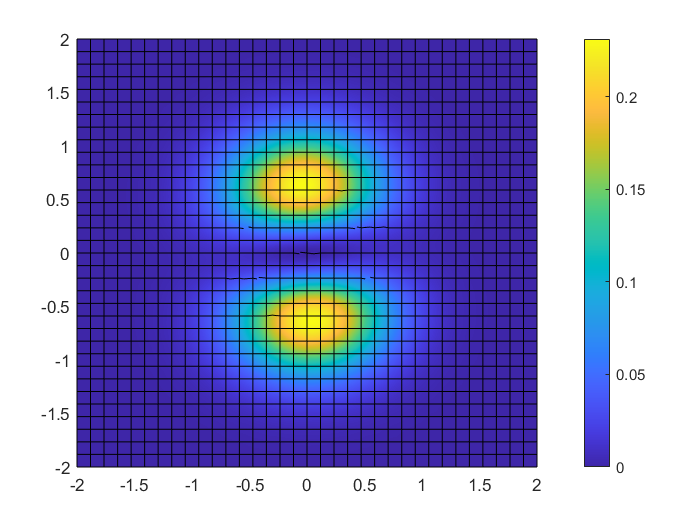}
\put(3,73){\textbf{(b)}}
\end{overpic}
\label{7b}}
\subfigure{
\centering
\begin{overpic}[scale=0.21]{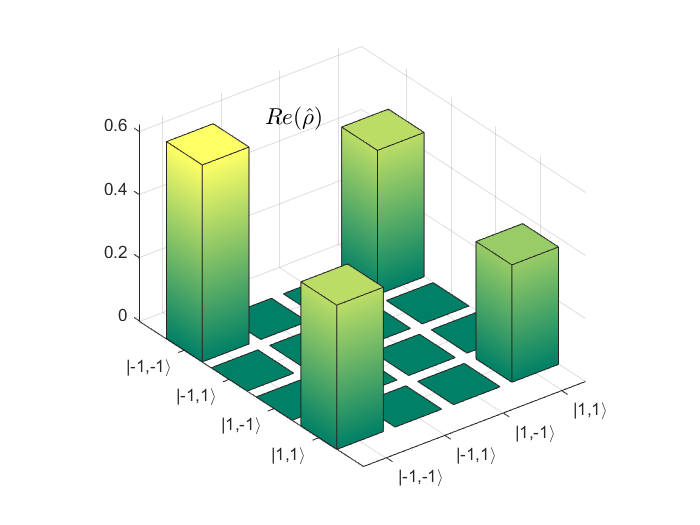}
\put(4,66){\textbf{(c)}}
\end{overpic}
\label{7c}}
\subfigure{
\centering
\begin{overpic}[scale=0.21]{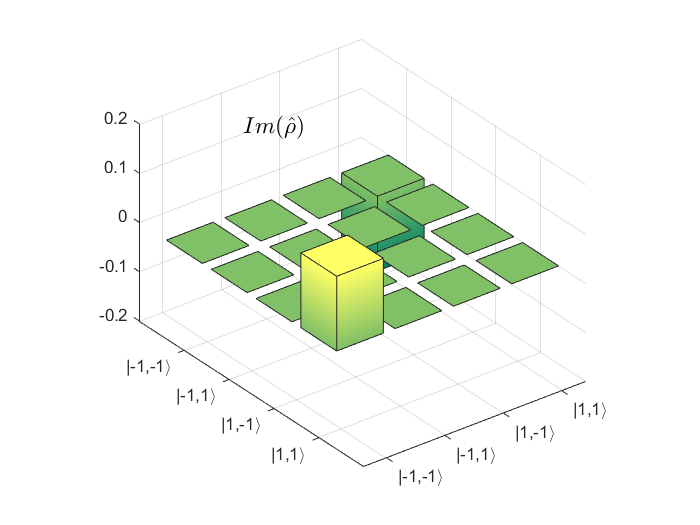}
\put(4,66){\textbf{(d)}}
\end{overpic}
\label{7d}}
\caption{The intensity distribution of the light beam after the first(a) and second(b) projections. (c) and (d) represents the real and imaginary parts of the OAM condition after reconstruction, respectively.}
\label{7}
\end{figure}
In the OAM scenario, we can calculate that the fidelity after the reconstruction reaches around $98\%$ compared to that of before the incident beam. Considering Figs.\ref{6}-\ref{7}, the quantum bit configuration also performs well in the optical fiber. But the situation of quantum distribution is not discussed in this paper, and we will study it in our future work.\\
\section{Conclusion}
In this paper, we propose a method of quantum state modulation that allows us to obtain quantum states with higher fidelity in different situations. We consider the performance in optical fibers and find that the tuned quantum states exhibit better stability, in which high fidelity can be achieved at $\alpha=0.8$, which ensures that our tuned states can be transported with better stability and a wider range. We develop the discussion in the APD  (amplitude and phase damping) channel as well as in the fiber condition. The results show that the tuned entangled state in the APD channel performs the regular state to a certain extent, and its corresponding fidelity improves by $7\%$ at $10\mu s$, both for the 4 qubits case and the 6 qubits case, with the advantage becoming more pronounced with increasing time. The fidelity improvement is also more significant when considering only the OAM entangled state, which at $100\mu s$ improves the fidelity by about $10\%$. 
 Compared with the conventional situation, our entangled states can generate a better fidelity and obtain a better stability.The main scope of our work is to improve the transmission fidelity of OAM states in fiber. \\
 
\textbf{Author contributions}: H. Zhang performed the
simulation, supervised by C. Zhao. H. Zhang and C. Zhao prepared the manuscript. C. Zhao supervised the project.

\textbf{Disclosures}\\
Others declare no conflicts of interest.

\textbf{Data Availability Statement}\\
Data and simulation codes related to
this work are available from the corresponding author upon reasonable request.

\bibliography{main}

\end{document}